# Instability of $q$-expectation value

S. ABE[1,2,3]

[1]*Department of Physical Engineering, Mie University, Mie 514-8507, Japan*[*)]

[2]*Institut Supérieur des Matériaux et Mécaniques Avancés, 44 F. A. Bartholdi, 72000 Le Mans, France*

[3]*Inspire Institute Inc., McLean, Virginia 22101, USA*

**Abstract** — $q$-Expectation value of a physical quantity is widely used in nonextensive statistical mechanics. Here, it is shown that the $q$-expectation value is not stable under small deformations of a probability distribution function, in general, whereas the ordinary expectation value is always stable.

___________________________

[*)] Permanent address

Given a statistical mechanical system, perform a measurement to obtain a probability distribution $\{p_i\}_{i=1,2,...,W}$, where $W$ is the number of microstates and is very large ($2^{10^{23}}$, typically). Perform a measurement again on the same system prepared in the same state as before. Then, another probability distribution $\{p'_i\}_{i=1,2,...,W}$ will be obtained. If the measurements are physically meaningful, then the difference between these two probability distributions is small. One naive way to quantify such a difference may be to compare the expectation values of a certain physical quantity, say $Q = \{Q_i\}_{i=1,2,...,W}$, with respect to those two probability distributions. And, it seems quite legitimate to anticipate that these expectation values are also close to each other.

In this article, we show that this naive anticipation is not guarantied, in general, and needs justification at a certain level. More precisely, we require that the following predicate should hold for *any* two probability distributions, $\{p_i\}_{i=1,2,...,W}$ and $\{p'_i\}_{i=1,2,...,W}$:

$$(\forall \varepsilon > 0) \ (\exists \delta > 0) \ \left( \|p - p'\|_1 < \delta \ \Rightarrow \ |\langle Q \rangle - \langle Q \rangle'| < \varepsilon \right), \tag{1}$$

where $\|p - p'\|_1 = \sum_{i=1}^{W} |p_i - p'_i|$ is the $l^1$ norm defined in the space of probability distributions and $\langle Q \rangle$ ($\langle Q \rangle'$) denotes the expectation value of $Q$ with respect to $\{p_i\}_{i=1,2,...,W}$ ($\{p'_i\}_{i=1,2,...,W}$). One might consider other kinds of norms, but it turns out that what is relevant to our discussion is the $l^1$ norm. The definition of expectation value $\langle Q \rangle$ is said to be "stable", if the condition in Eq. (1) is satisfied for *any* two probability distributions, $\{p_i\}_{i=1, 2, ..., W}$ and $\{p'_i\}_{i=1, 2, ..., W}$. This concept is somewhat analogous to Lesche's stability condition on entropic functionals [1] (see also Refs. [2-6]).

We are going to consider two different definitions of expectation value. One is the ordinary expectation value

$$\langle Q \rangle_1 = \sum_{i=1}^{W} Q_i \, p_i \tag{2}$$

and the other is the so-called $q$-expectation value

$$\langle Q \rangle_q = \sum_{i=1}^{W} Q_i \, P_i^{(q)}, \tag{3}$$

where $\{P_i^{(q)}\}_{i=1, 2, ..., W}$ is the escort probability distribution associated with $\{p_i\}_{i=1, 2, ..., W}$ [7]:

$$P_i^{(q)} = \frac{(p_i)^q}{\sum_{j=1}^{W}(p_j)^q}. \tag{4}$$

The latter might give the reader an exotic impression, but actually it is widely used in the field of nonextensive statistical mechanics [8-10]. Clearly, $\{p_i\}_{i=1, 2, ..., W}$ and $\{P_i^{(q)}\}_{i=1, 2, ..., W}$ coincide with each other at least in two cases: the equiprobability distribution, $p_i = 1/W = P_i^{(q)}$ $(i = 1, 2, ..., W)$, or $q = 1$.

It is noticed that both Eqs. (2) and (3) are assumed to be finite. One might claim that there is a class of probability distributions that has divergent moments, such as power-law probability distributions. However, $W$ is large but still finite. The limit $W \to \infty$ is evaluated at the last stage of examining Eq. (1). Also, one should recall the fact that any experimentally obtained distribution never has an infinite support.

Let us show that the ordinary expectation value in Eq. (2) satisfies the condition in Eq. (1). For this purpose, we evaluate the quantity $\left|\langle Q \rangle_1 - \langle Q \rangle_1'\right|$ as follows:

$$\left|\langle Q \rangle_1 - \langle Q \rangle_1'\right| = \left|\sum_{i=1}^{W} Q_i (p_i - p'_i)\right|$$

$$\leq \sum_{i=1}^{W} |Q_i| |p_i - p'_i|$$

$$\leq |Q|_{\max} \cdot \|p - p'\|_1, \qquad (5)$$

where $|Q|_{\max} \equiv \max\{|Q_i|\}_{i=1, 2, ..., W}$. This means that there exists $\delta$ such that $\delta = \varepsilon / |Q|_{\max}$. Thus, the definition of the ordinary expectation value is stable.

The situation is, however, completely different for the $q$-expectation value in Eq. (3): the condition (1) does not hold for $\langle Q \rangle_q$ unless $q = 1$, in general. To see it, let us consider the following infinitesimal deformations considered in Ref. [1]:

(i) for $0 < q < 1$;

$$p_i = \delta_{i1}, \quad p'_i = \left(1 - \frac{\delta}{2}\frac{W}{W-1}\right)p_i + \frac{\delta}{2}\frac{1}{W-1}, \tag{6}$$

(ii) for $q > 1$;

$$p_i = \frac{1}{W-1}(1 - \delta_{i1}), \quad p'_i = \left(1 - \frac{\delta}{2}\right)p_i + \frac{\delta}{2}\delta_{i1}. \tag{7}$$

In both cases, the $l^1$ norms are

$$\|p - p'\|_1 = \delta. \tag{8}$$

The independence of the norm from $W$ is an important point for examining stability freely from the system size. In the case (i), we have

$$\sum_{i=1}^{W} (p_i)^q = 1, \quad \sum_{i=1}^{W} (p'_i)^q = \left(1 - \frac{\delta}{2}\right)^q + \left(\frac{\delta}{2}\right)^q (W-1)^{1-q}, \tag{9}$$

and

$$\sum_{i=1}^{W} (p_i)^q = (W-1)^{1-q}, \quad \sum_{i=1}^{W} (p'_i)^q = \left(\frac{\delta}{2}\right)^q + \left(1 - \frac{\delta}{2}\right)^q (W-1)^{1-q}, \tag{10}$$

in the case (ii).

Now, let us examine stability of the $q$-expectation value in Eq. (3) under these deformations. From Eqs. (6), (7), (9), and (10), we find that, in the case (i),

$$\left|\langle Q\rangle_q - \langle Q\rangle'_q\right| = \left|Q_1 - \frac{(1-\delta/2)^q Q_1 + \left(\delta/[2(W-1)]\right)^q \sum_{i=2}^{W} Q_i}{(1-\delta/2)^q + (\delta/2)^q (W-1)^{1-q}}\right|$$

$$\to \left|\overline{Q} - Q_1\right| \quad (W\to\infty), \tag{11}$$

and, in the case (ii),

$$\left|\langle Q\rangle_q - \langle Q\rangle'_q\right| = \left|\frac{W\overline{Q} - Q_1}{W-1} - \frac{(\delta/2)^q Q_1 + [(1-\delta/2)/(W-1)]^q \left(W\overline{Q} - Q_1\right)}{(\delta/2)^q + (1-\delta/2)^q (W-1)^{1-q}}\right|$$

$$\to \left|\overline{Q} - Q_1\right| \quad (W\to\infty), \tag{12}$$

where $\overline{Q} = (1/W)\sum_{i=1}^{W} Q_i$ is the arithmetic mean associated with the equiprobability distribution. Therefore, the condition in Eq. (1) is not satisfied by the *q*-expectation value.

In conclusion, we have shown that the *q*-expectation value is not a uniformly continuous functional and is unstable under small deformations of the underlying probability distribution, in general. It is also of interest to recognize that the limits $W\to\infty$ and $q\to 1$ do not commute. The *q*-expectation value may be a useful mathematical tool for characterizing a special class of probability distributions such as asymptotically power-law distributions [11], but it does not coincide with the physical one that an experimentalist obtains from his data in the usual manner [12]. Thus, the present result seems to have an important implication for the foundations of nonextensive statistical mechanics. It suggests that the correct definition to be employed in nonextensive statistical mechanics may not be the *q*-expectation value. Accordingly, the ordinary-expectation-value formalism should be reexamined. Nonextensive statistical mechanics with the ordinary expectation value was considered in Ref. [13] in an incomplete manner, and its complete formulation was presented in Ref. [14]. This point will be discussed in wider perspective elsewhere.

* * *

The discussions with T. Biro, R. Hanel, S. Thurner, C. Tsallis, and G. Wilk are gratefully acknowledged. This work was supported in part by a Grant-in-Aid for Scientific Research from the Japan Society for the Promotion of Science.